\documentclass[aps, prl, twocolumn, superscriptaddress, showpacs, byrevtex, floatfix]{revtex4}
\usepackage{graphicx}

\begin{document}
\title{Correlating Quasi--Electron
Relaxation with  
 Quantum Femtosecond Magnetism in the 
Order Parameter
Dynamics of Insulating Manganites}
\author{T. Li}
\affiliation{Ames Laboratory and
	Department of Physics and Astronomy, Iowa State University, Ames,
	Iowa 50011, U.S.A.}

\author{A. Patz}
\affiliation{Ames Laboratory and
	Department of Physics and Astronomy, Iowa State University, Ames,
	Iowa 50011, U.S.A.}

\author{P. Lingos} 
\affiliation{Department of Physics, University of Crete, 
	Box 2208,  Heraklion, Crete, 
	71003, Greece}

\author{L. Mouchliadis} 
\affiliation{Department of Physics, University of Crete, 
	Box 2208,  Heraklion, Crete, 
	71003, Greece}

\author{L. Li}
\affiliation{Department of Materials Science and Engineering,  University of Tennessee, Knoxville, Tennessee 37996, U.S.A}

\author{J. Yan}
\affiliation{Department of Materials Science and Engineering,  University of Tennessee, Knoxville, Tennessee 37996, U.S.A}

\author{I. E. Perakis}
\affiliation{Department of Physics, University of Crete, 
	Box 2208, 
	Heraklion, Crete, 
	71003, Greece}

\affiliation{ Institute of Electronic Structure \& Laser, Foundation for
	Research and Technology--Hellas, 
	Vassilika Vouton, 71110 Heraklion, Crete, Greece}

\author{J. Wang}
\affiliation{Ames Laboratory and Department of Physics and
	Astronomy, Iowa State University, Ames, Iowa 50011, U.S.A.}

\date{\today}

\begin{abstract}
Femtosecond (fs)--resolved simultaneous measurements of charge and spin dynamics reveal the coexistence of two different
 quasi--particle excitations in colossal magneto-resistive
(CMR)  manganites, with
 {\em fs} and {\em ps} relaxation times respectively.
Their populations reverse size 
above a {\em photoexcitation--intensity--threshold} coinciding 
with a ``sudden'' antiferro--to--ferromagnetic 
 switching during 
 $<$100 fs laser pulses. 
We present evidence that fast, metallic, mobile quasi--electrons 
dressed by {\em quantum spin fluctuations} coexist with 
 slow, localized, polaronic charge carriers in 
non--equilibrium phases.
This may be central to CMR transition and leads to a laser--driven charge reorganization simultaneously with  
quantum fs 
magnetism via an emergent 
quantum--spin/charge/lattice transient coupling.
\end{abstract}
\pacs{78.67.Wj, 73.22.Pr, 78.47.J-,78.45.+h}
\maketitle

Traditionally, quantum material phases are tuned 
by static 
parameters such as chemical dopants, 
pressure, or magnetic fields. {\em Spontaneous 
coherence} induced in this way, 
e.g. between many--body 
states separated by the Mott--Hubbard insulator gap,  
can  establish new orders via  
equilibrium phase transitions. 
{\em Non--equilibrium} phase transitions 
may  be similarly triggered 
 by  non--local, time--dependent electron--hole ($e$--$h$) 
coherence 
driven by a  fs laser pulse \cite{Chemla,Axt98}. 
Due to the ``sudden'' 
time--dependent 
change in the Hamiltonian, 
the equilibrium state is no longer the 
ground state  of
the coupled light--matter system, 
which creates a quasi--instantaneous 
initial condition for 
time evolution
 of material phases.
Strongly--correlated 
 states,
determined by many--electron ordering and coherence 
arising, e.g., when 
 local interactions
exceed or compare to the kinetic energy,
are particularly responsive to
such non--adiabatic excitations.  
In contrast, fs excitations merely perturb the 
{\em ``rigid"} 
electronic bands of 
weakly--correlated materials 
(e.g. semiconductors). 
In the manganites, laser--driven 
 bonding mediated by
quantum spin flip/canting fluctuations was shown to induce a magnetic phase 
transition during
 $<$100fs  pulses \cite{Li-2013}.
The {\em quantum 
femtosecond magnetism}  originates from 
transient modification
of inter--atomic {\em e}--hopping amplitude 
by the 
laser {\em E}--field \cite{Li-2013}, which {\em non--adiabatically} generates spin--exchange coupling and ferromagnetic correlation,  
as
photoelectron hopping  
{\em simultaneously} flips local spins. 

Complex materials 
such as manganites 
involve  simultaneous ordering
of multiple  degrees 
of freedom: spin, lattice, charge/orbital orders, etc \cite{SheuPRX,IchikawaNatMat,FiebigSci,PolliNat,RiniNat}.
The elementary excitations then depend on a complex 
set of coupled order parameters 
with large fluctuations which makes it difficult to underpin
their microscopic compositions   
\cite{Milward,Aaron}.
Although strong coupling of electronic, magnetic, and lattice 
degrees of freedom in the manganites is known to lead to 
 coexisting  
insulating/lattice--distorted/antiferromagnetic (AFM) 
and metallic/undistorted/ferromagnetic (FM) 
regions
of sizes $\sim$10--300nm 
\cite{Dagotto}, 
 the relevant 
quasi--particles remain controversial.  
While electrons
localized by Jahn--Teller 
(JT) lattice distortions   
dominate the AFM insulating
 state,
some theoretical studies have proposed that 
the sensitivity to small perturbations leading 
to 
CMR phase transition to 
a FM metallic state
is due to coexisting 
mobile {\em minority} electrons mediated by 
classical spin canting
\cite{Krish,Ramakrish,Milward}. 
Moreover, while the strong spin--charge coupling should considerably correlate the corresponding {\em fs} dynamics, the photoexcitation--threshold observed for fs spin generation was absent in the measured optical conductivity \cite{Li-2013, Okimoto} and the exact linkage between the two is still elusive.  
The  simultaneous probing 
of fs spin \cite{Li-2013,Bigot,Subkhangulov}  and charge \cite{PRB}  
dynamics
may
dynamically disentangle 
 degrees of freedom coupled in equilibrium 
and reveal crucial 
 many--body mechanisms. 

This Letter uses fs pump--probe 
spectroscopy to identify the 
quasi--particle excitations 
of Pr$_{0.7}$Ca$_{0.3}$MnO$_3$ (PCMO) manganites
and quantify 
their  
dynamical properties.
For this  
we excite  non--equilibrium electron 
populations close to the insulator
 energy gap and 
then 
 probe with fs resolution 
their effects on  both
 differential reflectivity 
and magneto--optical responses 
at 1.55eV and 3.1eV. 
 When 
pump/probe 
are {\em both} 
tuned at 
1.55 eV, we observe  
a two--step bi--exponential relaxation of charge excitations
absent at 3.1eV.
These two components, 
characterized by  distinct fast fs ($\tau^{\mathrm{fs}}$) and 
slow ps ($\tau^{\mathrm{ps}}$)
relaxation times, 
disappear  at higher temperatures.
Intriguingly, the ratio of their  amplitudes 
displays a
{\em  pump-fluence-threshold} nonlinear dependence 
that {\em coincides} 
with the threshold 
for  {\em fs} AFM$\rightarrow$FM switching. 
We present calculations indicating 
{\em coexistence 
 in a
non--equilibrium phase} 
 of 
 fast, mobile, metallic  
 quasi--electrons dressed by {\em quantum 
spin fluctuations}
($\tau^{\mathrm{fs}}$)
 with  slow, localized polaronic
 carriers
($\tau^{\mathrm{ps}}$).
The laser--induced 
rearrangement 
of these majority and minority carriers
creates a critical 
 non--thermal 
population 
of 
quasi--electrons with strongly--coupled 
spin--charge degrees of freedom, which   
drives a simultaneous
AFM$\rightarrow$FM switching 
via quantum spin--charge--lattice 
dynamical coupling.

\begin{figure}[floatfix]
\begin{center}
\includegraphics[scale=0.35]{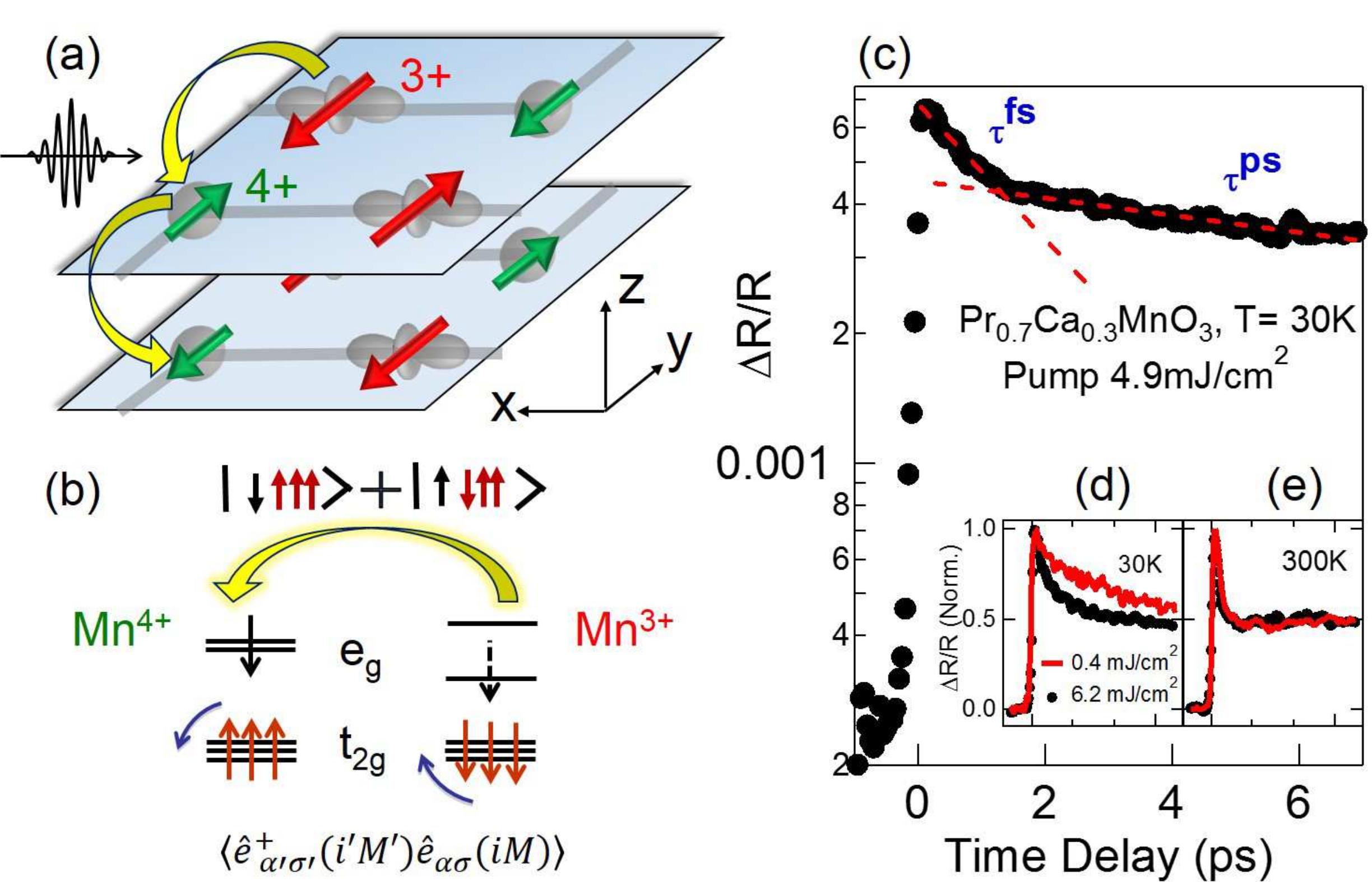} 
\caption{(Color online) Illustration of (a) 
fs $e$--$h$ excitations in
CE--AFM/CO/OO ordered manganites and (b) laser--driven 
off-diagonal bonding--order, via quantum 
spin canting, 
of composite fermion
quasi--particles (supplementary section).
 (c): fs--resolved $\Delta$R/R charge dynamics 
for 1.55eV pump/probe excitation, plotted on a log--scale.
Dashed lines highlight two distinct components of  
bi--exponential decay. (d)-(e): 
Comparison of normalized 
$\Delta$R/R  for two pump fluences marked at (d) 30K and (e) 300K.}
\end{center}
\label{Fig1}
\end{figure}

We consider the 
CE--AFM--insulator 
state characterized by coexisting 
charge  (CO),
orbital  (OO), 
and magnetic orders
 \cite{Tokura,Dagotto}. 
Here, AFM--coupled 
charge--modulated 
zig-zag chains
 have 
alternating Mn$^{3+}$/Mn$^{4+}$ ions (CO), 
FM spins,   
and  JT--distorted lattice sites 
with populated  orbitals 
pointing along
the chain (OO)
(Fig. 1(a)).
The JT interaction of a
 localized $e_g$--electron with its
surrounding  Mn$^{3+}$O$_{6}$ octahedron 
 splits the two degenerate Mn$^{3+}$ states and 
results in a {\em polaronic insulator} with 
populated lower level
(JT energy gain $E_{JT}$) \cite{Krish,Tokura}.  
We study here Pr$_{0.7}$Ca$_{0.3}$MnO$_3$  single-crystals 
 grown by 
the floating--zone method. {\em All equilibrium phases
 are  insulating}, 
with CO/OO order  below $\sim$200K and CE--AFM order 
below $\sim$140K. 
A Ti:Sapphire amplifier laser 
beam, 
with pulse duration of 35fs and repetition rate of 1kHz, was used 
in 
{\em fs} pump--probe spectroscopy 
measurements of 
 differential reflectivity $\Delta R/R$, 
 magneto-optical Kerr effect (MOKE, $\Delta\theta_k$), and 
magnetic circular dichroism (MCD, $\Delta\eta_k$).
We thus trace the fs spin and charge dynamics for 
magnetic field B$\leq$ 0.25T \cite{note1}.

$e_g$--electron charge fluctuations are 
restricted  by  exchange interaction with 
the  local $S$=3/2 spins 
formed by filling all three t$_{2g}$--orbitals 
\cite{Tokura,anderson}
and by 
suppression of double--occupancies (Mn$^{2+}$) 
by the 
strong 
local interactions.
In classical--spin thermodynamic
scenarios, it is energetically favorable for 
$e_g$--electrons (spin $s$=1/2) 
to move within a single chain 
so that the
spins remain FM--coupled 
via strong Hund's rule interaction $J_H {\bf S}_{i}\cdot 
{\bf s}_{i}$. 
For quantum spins, however,  photoelectrons can also hop to 
sites with anti--parallel $t_{2g}$ spins without magnetic 
energy cost,
illustrated in 
Fig. 1(b). 
This is possible  
by forming quantum states 
with 
the same total spin 
$J$
after flipping $t_{2g}$--spins via 
$J_H S^{\pm }_{i}\cdot s^{\mp }_{i}$ 
electron--magnon
coupling that leads to quantum spin canting \cite{Kapet-corr} (supplementary section). 
These ultrafast fluctuations mediate non-local off-diagonal inter-atomic bonding (Fig. 1(b)) 
and dynamically--entangle neigboring 
AFM chains, Fig. 1(a),
which competes with the AFM surroundings 
to establish a 
metastable 
state ({\em quantum femtosecond  magnetism} 
\cite{Li-2013}). 
Such  FM 
 correlation  
during the coherently-excited {\em e}-hopping 
has  no speed--limit
imposed by free energy or spin adiabaticity \cite{note,Krish,Mentink}.  

Fig.1(c) shows the typical {\em fs}--resolved 
$\Delta R(t)/R$ signal 
at 30K,  with {\em both}
pump/probe tuned at 1.55eV.
The 
non--equilibrium quasi--particles 
excited close to the
 insulator gap
with this relatively high pump fluence $\sim$4.9 mJ/cm$^{2}$
display  bi--exponential 
relaxation, 
with two distinct relaxation times 
$\tau^{\mathrm{fs}}$$\sim$530fs and $\tau^{\mathrm{ps}}$$\sim$5.7ps.
 This behavior is in striking contrast 
to 
the temporal  profiles of 
the low--fluence 
or high temperature 
signals (compare the  normalized 
$\Delta R/R$ traces in 
Figs. 1(d) and (e)). 
At 30K, the  $\tau^{\mathrm{fs}}$
 component is suppressed 
for smaller pump fluences of 
0.4 mJ/cm$^{2}$ (Fig. 1(d)). 
At
 300K, i.e. above the CO/OO  transition,
all pump fluences  give an
identical single--exponential decay, 
with   relaxation time 
much shorter than both  
$\tau^{\mathrm{fs}}$ and $\tau^{\mathrm{ps}}$ 
(Fig. 1(e)). 
Clearly the charge quasi--particle dynamics strongly depend on both photoexcitation 
intensity and ground--state order.

\begin{figure}[floatfix]
\begin{center}
\includegraphics [scale=0.37] {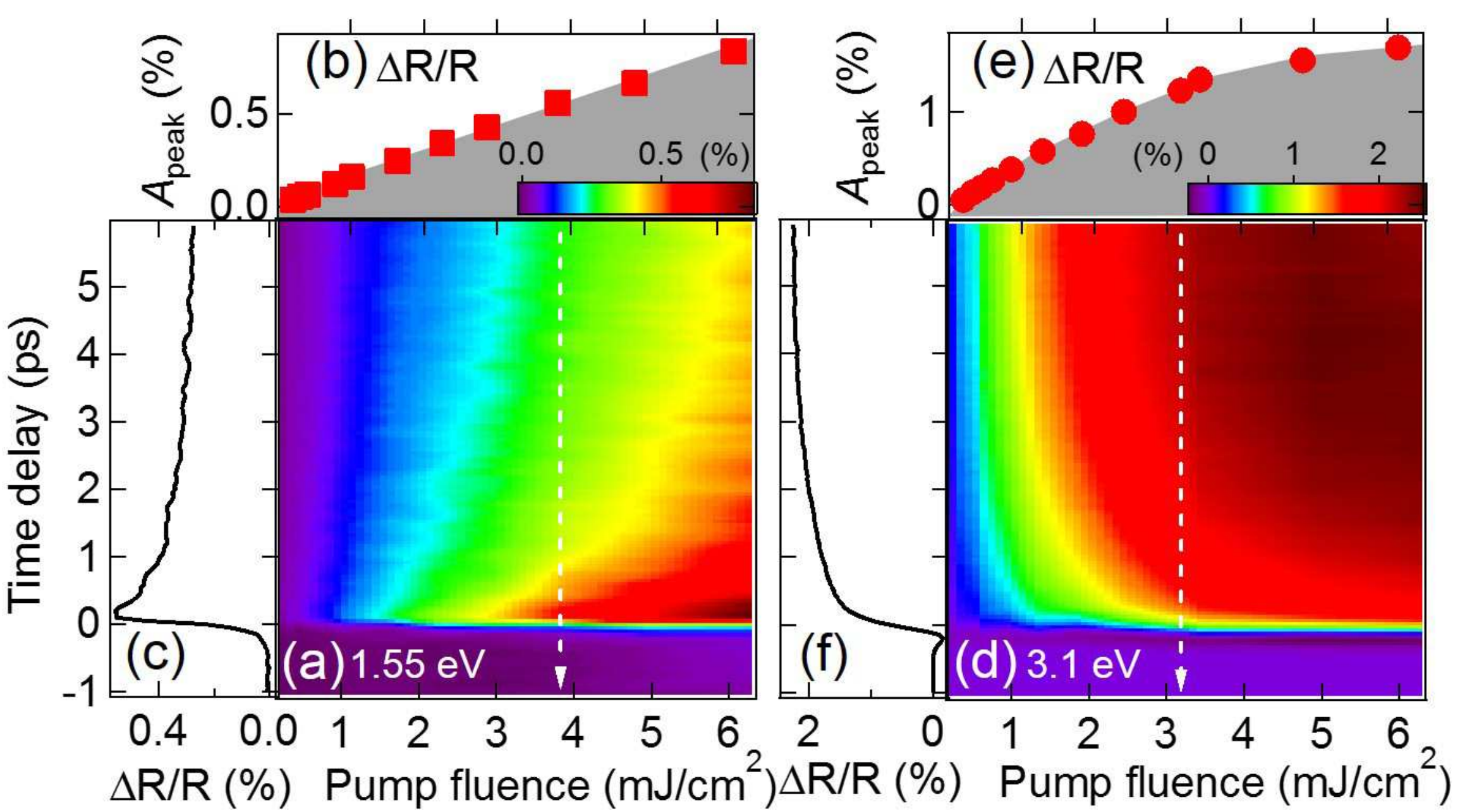}  
\caption{(Color online) (a)-(c): 
Ultrafast $\Delta$R/R dynamics 
under 1.55eV pump/probe photoexcitation. 
(a): 2D dependence on pump fluence and time delay at 30K; (b): peak amplitude as function of pump fluence; (c): temporal trace at 3.8mJ/cm$^2$ marked in (a).
(d)-(f): Same $\Delta$R/R plot as above, but
for  non--degenerate photoexcitation with 
1.55eV pump/3.1eV probe. }
\end{center}
\label{}
\end{figure}

Fig. 2 shows a  2D plot of $\Delta R/R$  as function
 of pump--fluence and probe time delay. The color gradients demonstrate 
distinct differences, along both axes, between probe frequencies that either 
couple directly  to [1.55 eV, Fig. 2(a)] or decouple from 
[3.1 eV, Fig. 2(d)] 
the insulator
 gap. 
While at 1.55eV the peak of
$\Delta R(t)/R$ 
 shows  almost linear fluence--dependence 
(Fig. 2(b)), at 3.1eV 
it displays  
nonlinear saturation 
(Fig. 2(e)).
For 1.55 eV pump/probe, 
$\Delta R(t)/R$ comes from 
 phase--space--filling by linearly--increasing
quasi--particle populations
near 
the  insulator  gap.
Its
temporal decay, 
Fig. 2(c), thus  reveals 
 two coexisting  quasi--particle 
populations, 
with relaxation times 
$\tau^{\mathrm{fs}}$ and  $\tau^{\mathrm{ps}}$ 
respectively. 
For 3.1eV probe/1.55eV pump, 
$\Delta R(t)/R$ 
reflects  a {\em fs} increase and  saturation of
the conductivity, with 
spectral--weight transfer to low energies 
due to a fs pump--induced 
decrease
in the 
insulator gap.
Fig. 2(f) shows 
a further 
{\em ps} increase 
of $\Delta R(t)/R$, which  
reflects a slower  phonon--related conductivity increase.

We now compare this charge relaxation to the spin dynamics extracted 
 from the 
fs--resolved 
magnetic signals. 
Fig. 3(c) clearly shows threshold
for {\em fs} spin photogeneration above a critical pump--fluence of 2-3mJ/cm$^2$ where both MOKE and MCD show  the {\em same large 
quasi--instantaneous jump} (inset). 
Despite this, Figs. 2(b) 
and 2(e) show  smooth 
{\em thresholdless}
fluence--dependence 
of $\Delta R/R$.
Disentangling 
the $\tau^{\mathrm{fs}}$ and  $\tau^{\mathrm{ps}}$ 
components of $\Delta R(t)/R$ provides the missing link between spin and charge 
quantum excitations. 
Fig. 3(a) shows the pump--fluence--dependences of the 
amplitudes  
A$^{fs}$ and A$^{ps}$, via bi--exponential fit, and their sum
 A$^{sum}$=A$^{fs}$+A$^{ps}$
 (inset). 
While A$^{sum}$ appears  linear, 
the two populations  A$^{fs}$ and A$^{ps}$ 
reverse their 
magnitudes 
 with increasing excitation
(Fig. 3(a)).
Most intriguingly, a threshold increase of the short--lived ($\tau^{\mathrm{fs}}$) minority population 
is seen in Fig. 3(b) by plotting the fraction
 F=A$^{fs}$/A$^{sum}$. This apparent threshold 
{\em coincides} 
 with the threshold for {\em fs} spin generation 
in Fig. 3(c), while 
$\tau^{\mathrm{fs}}$ 
and $\tau^{\mathrm{ps}}$ 
times remain fairly constant 
(inset of
Fig. 3(b)).
This
direct correlation of AFM$\rightarrow$FM switching  
with critical increasing the proportion
of the minority $\tau^{\mathrm{fs}}$ population
suggests the emergence of a 
quasi--particle 
excitation
composed of  
strongly--coupled spin and charge degrees of 
freedom.

\begin{figure}[floatfix]
\begin{center}
\includegraphics [scale=0.4] {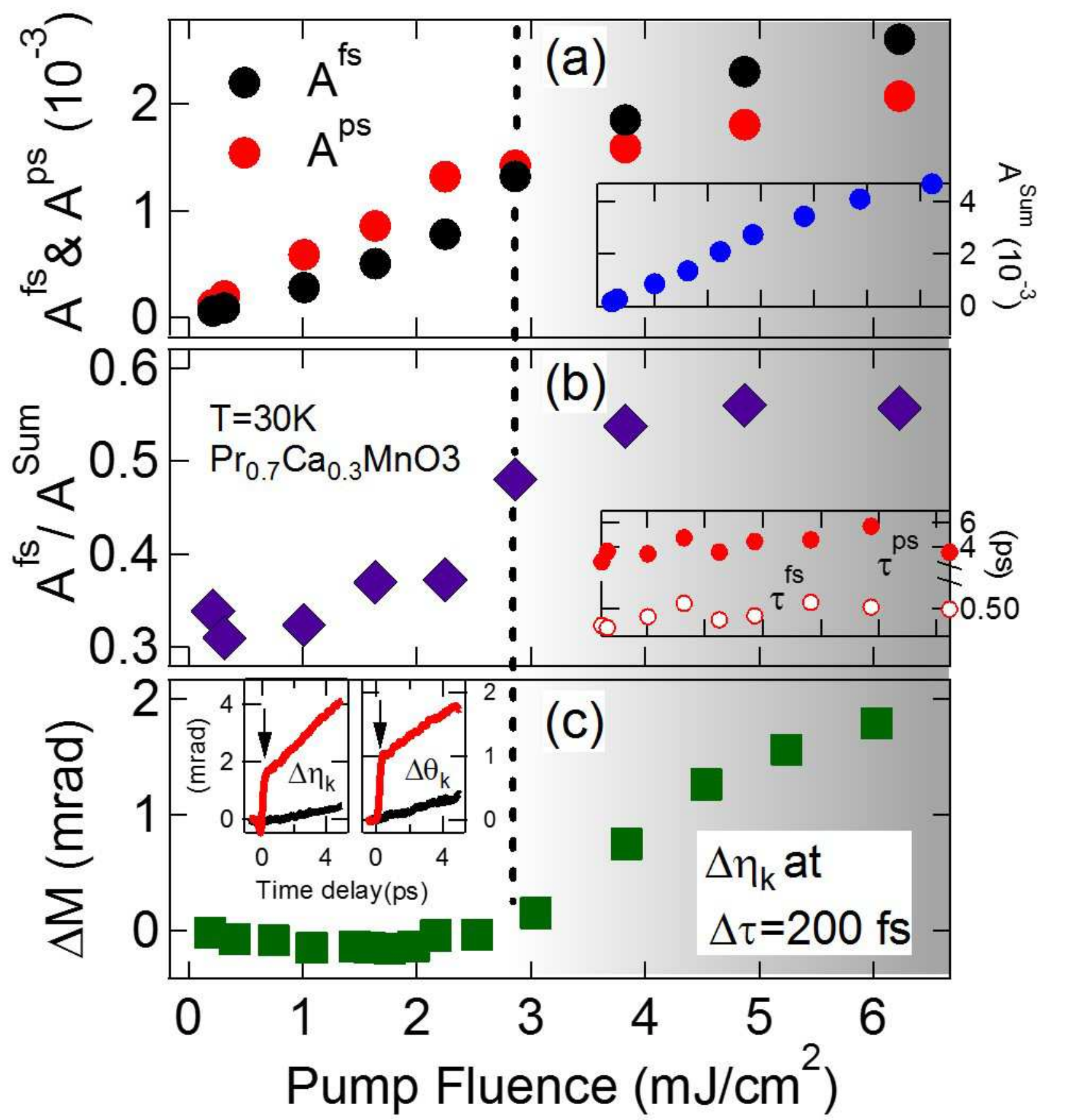}  
\caption{(Color online) Photoexcitation dependence. 
(a): Amplitudes of fast component 
 A$_{fs}$ 
(black dots), slow component A$_{ps}$ (red dots),
 and 
 A$^{sum}$= A$_{fs}$+ A$_{ps}$   
 (inset). (b): 
Fraction
 $F$=A$^{fs}$/A$^{sum}$ 
(blue rhombus) and the two distinct relaxation times (inset). 
(c): Photoinduced fs magnetization  $\Delta M$ 
extracted from $ \Delta\eta_k$
at 200fs 
(green rectangle).
Inset: 
 $\Delta\eta_k$ and $\Delta\theta_k$ dynamics 
for 5.6mJ/cm$^2$ (red) and 0.8mJ/cm$^2$ (black). 
show the same  ``sudden'' magnetization (arrow). 
All error bars within the markers. }
\end{center}
\label{}
\end{figure}

To explore this issue, we model  the non--adiabatic 
\cite{note}
spin--charge  quantum 
correlation
that dresses  $e$--$h$ excitations
{\em during} 
the fs 
timescales of coherent 
light--matter coupling. 
For this, we solve  the quantum--kinetic 
equations of motion 
of the spin--dependent  density matrix 
that describes  spin/charge non--equilibrium
 populations and 
inter--site coherences involving  atomic many--body states 
(supplementary section).
In the ground state
(Fig. 1(a)),   
fully--localized JT--polarons (majority carriers)
gain lattice energy $E_{JT}$ 
by populating
alternating 
Mn$^{3+}$ sites (site 1 in Fig.4(a)),   
with total spin 
$J$=$S$+1/2 and 
parallel $e_g$ and 
$t_{2g}$ spins.
In the  
deep--insulating limit  
of large $E_{JT}$ \cite{Krish}, we  
neglect 
electron hopping
along  FM chains, which  does not change the total spin.
We focus on  
quantum correlations  
between two 
neigboring  AFM  
atoms in different chains 
 (yellow arrows, Fig. 1(b)), 
driven by the laser 
E--field 
with central frequency 
$\hbar \omega_p$$\sim$$E_{JT}$.
The strong charge fluctuations 
during this fs pump pulse 
involve  hopping  of 
the $e_g$--electron
from the Mn$^{3+}$ atom 
($\varepsilon$=$-E_{JT}$)
to the 
JT--undistorted 
Mn$^{4+}$ atom ($\varepsilon$=0) with 
 {\em anti--parallel}  $t_{2g}$ spin
$S_z$=-$S$
(site 2 in Fig.4(b)). Such 
 virtual \cite{Ramakrish} 
and laser--driven 
fluctuations 
across the JT gap 
are faster than the 
JT distortions 
for hopping amplitudes
$t_0$$\gg \hbar \omega_{ph}$ \cite{Krish}, so 
 for now our simulation ignores 
JT displacements (phonon frequency $\omega_{ph}$) to examine the roles of quantum charge/spin fluctuations. 

Fig.4 shows all non--equilibrium 
 spin--resolved 
 populations 
of the two above-discussed sites 
(Figs. 4(a) and 4(b))
 and the z--component of the total $t_{2g}$--spin,
$S_z$=$S_z$(1)+$S_z$(2)
(Fig. 4(c)). 
Here $J_H$$\rightarrow$$\infty$,  
so an electron can hop 
between  AFM sites 
{\em only} by simultaneously flipping t$_{2g}$  spins \cite{anderson}. 
This results in 
correlated 
spin--charge non--adiabatic 
 dynamics. 
The bottom panel of 
 Fig.4(a) shows the photoexcited
 hole 
population
 (Mn$^{3+}$$\rightarrow$Mn$^{4+}$ excitation,
$J_z$=$S$+1/2$\rightarrow$$S_z$=$m$)
 of  the JT--distorted site 1. 
Excitation of  majority carriers 
does not change significantly 
the $m$=3/2 $t_{2g}$--spin. 
In contrast, the minority 
quasi--electrons
(Mn$^{4+}$$\rightarrow$Mn$^{3+}$ excitation, 
$S_z$=$-S$$\rightarrow$$J_z$=$M$)
populating site 2 (top  panel of 
Fig.4(b)) 
have mixed spin 
due to flipping of the opposite $e_g$ and $t_{2g}$ 
spins ($M$=$-S$+1/2). Such quasi--electron photoexcitations
thus induce a quantum dynamics of 
$S_z$,
which saturates 
with population inversion (Fig.4(c)). 
The fluence--dependence  of quasi--electron 
population 
then naturally
correlates   with that
of the fs spin, as in our experiment
 (Fig. 3(b)).
FM inter--chain correlation 
arises from this electron dressing 
by quantum 
spin fluctuations,    driven
by fs quantum--spin--canting in the AFM  
insulating state.

\begin{figure}[floatfix]
\begin{center}
\includegraphics [scale=0.35] {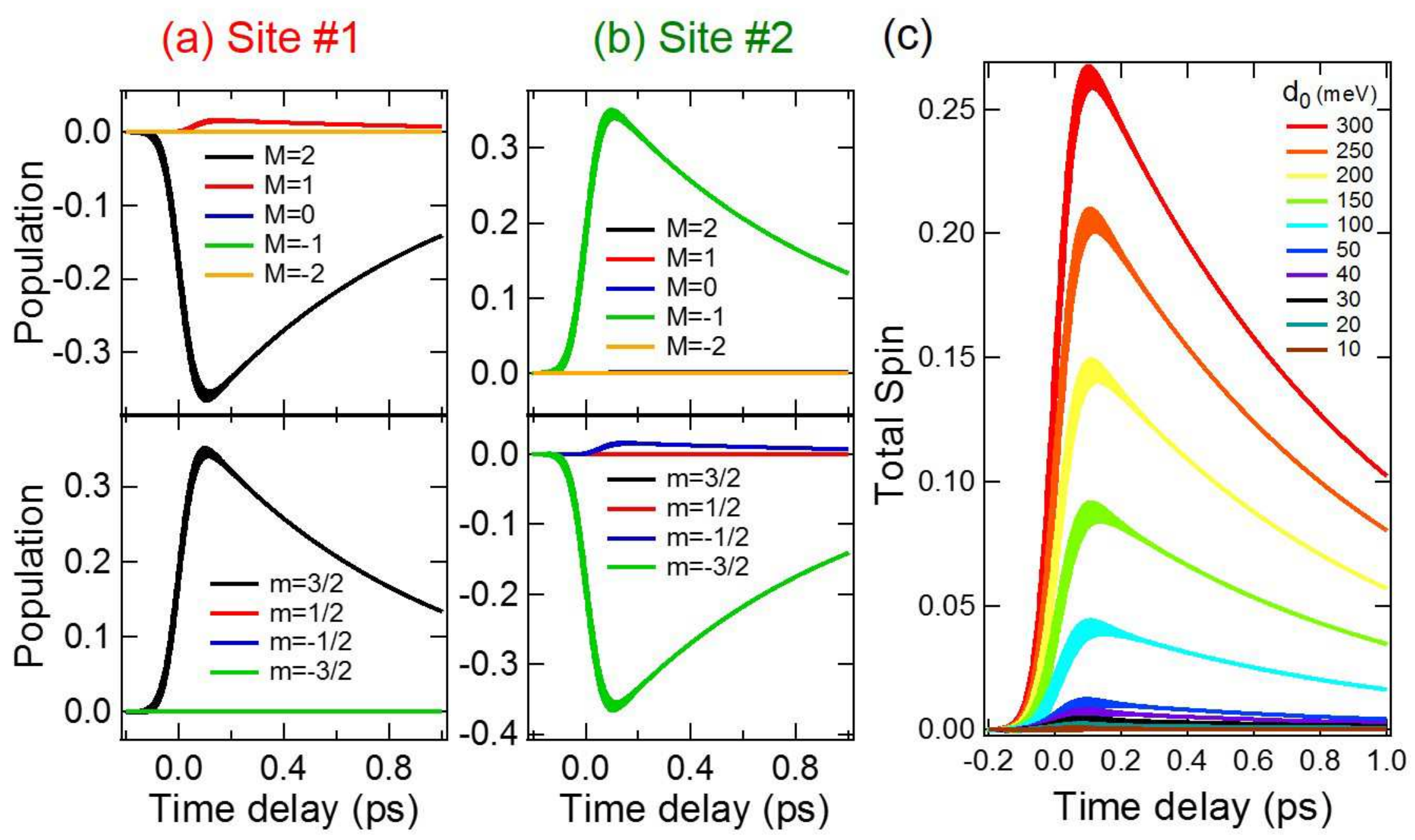}  
\caption{(Color online) 
Calculated time--dependence 
of   
(a): 
Spin--resolved 
Mn$^{3+}$ (upper panel) and Mn$^{4+}$ (lower panel) 
non--equilibrium 
populations, (b): 
Total $t_{2g}$--spin
under different Rabi energies $d_0$.
Here composite fermion 
populations, 
$T_1$=1ps, 
are riven 
by  $e$--$h$ photoexcitations
with lifetime 
$T_2$=50fs.
}
\end{center}
\label{}
\end{figure}

After 
photoexcitation, the
system is thereby left in an
excited 
state with  
 non--thermal  populations 
of two 
composite--fermion quasi--particles (supplementary section).
Subsequent  relaxation 
($\tau^{\mathrm{fs}}$ and $\tau^{\mathrm{ps}}$)
depends on the 
 quasi--particle 
energy dispersions,
shown in Fig. 5
along three  directions:
$k_x$ (along the chain), $k_y$ 
(perpendicular to the chain, along the same 
plane), and $k_z$ (perpendicular to the plane). We considered 
 one--electron 
excitations 
of the  CE--type CO/OO/AFM 
ordered periodic state \cite{unit} 
without spin--canting
(supplementary section). 
Fig. 5
 compares our quantum spin results 
to  the classical  limit 
$S$$\rightarrow$$\infty$,
 where we reproduce previous results 
\cite{Dagotto,Brink}.  
For classical spins,
an 
{\em adiabatic description} 
applies:
the diagonalized  electronic Hamiltonian
describes energy bands 
that depend on  {\em fixed}
local spin and JT--distortion patterns
 \cite{Krish,Dagotto}.  
For large $J_H$, 
the frozen CE--AFM spin pattern then 
only allows  photoelectron dispersion along a single 
FM chain 
\cite{anderson}. 
For quantum spins, however, 
photoelectrons move  by  {\em simultaneously} 
deforming 
 local spins
 (e.g. electron--magnon coupling \cite{Kapet-corr}).
They  decrease the 
insulator gap, Fig. 5, 
by hopping
{\em between} chains
parallel ($k_y$) or 
perpendicular ($k_z$) to the plane
(Fig. 1(a))  \cite{cant}.

\begin{figure}[floatfix]
\begin{center}
\includegraphics [scale=0.35] {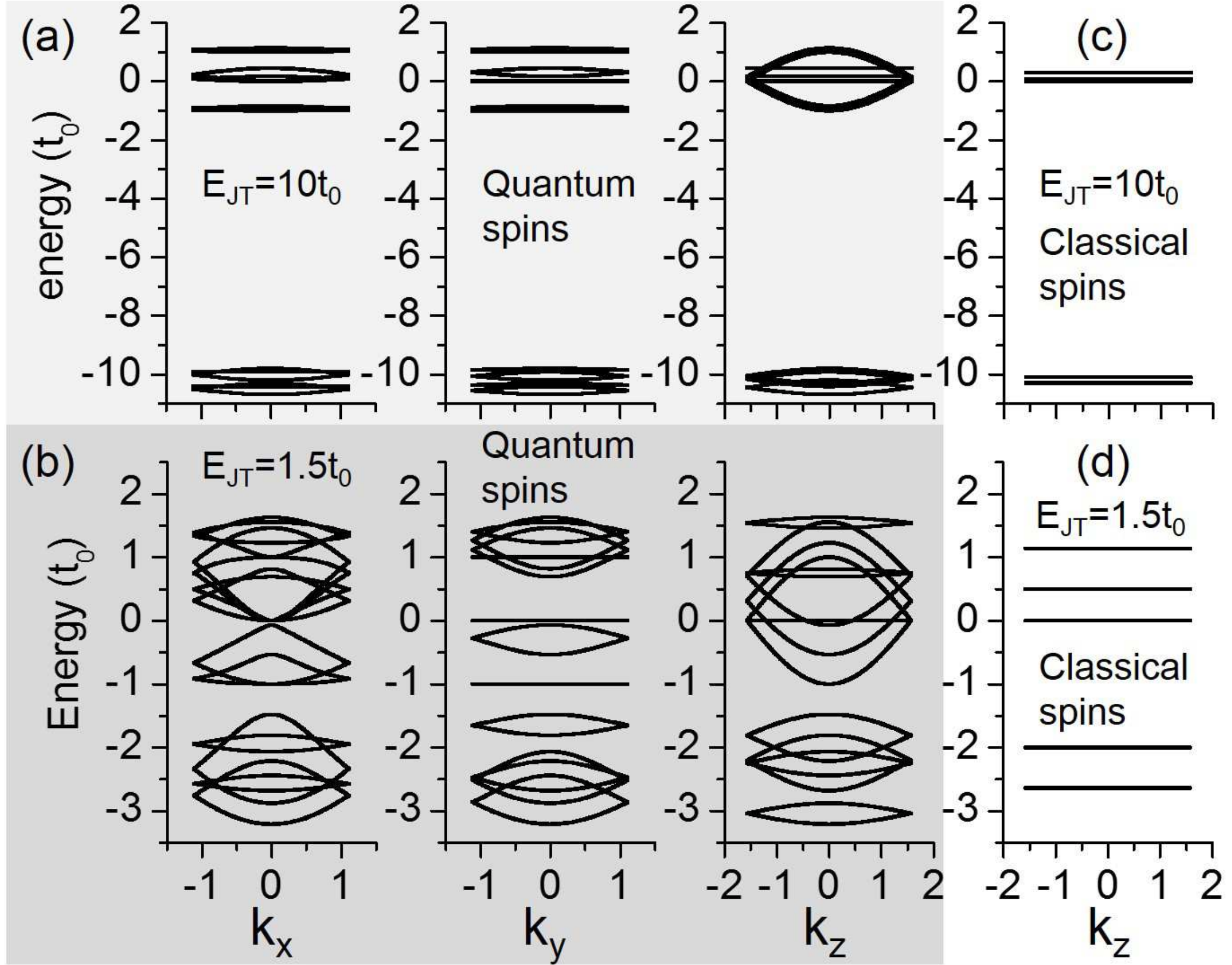}  
\caption{Calcualted composite--fermion 
energy dispersions. 
(a), (b): Quantum Spins,
(c), (d): Classical Spins (see text and supplementary section).
} 
\end{center}
\label{}
\end{figure}

For 
large 
 $E_{JT}$, 
Fig. 5(a) demonstrates anisotropic quasi--particle dispersions 
 with 
 energies  close to the 
Mn$^{3+}$ ($\varepsilon$=0) and Mn$^{4+}$ 
($\varepsilon$=$-E_{JT}$) 
localized levels.
CO  suppresses 
electron hopping 
{\em along the 
plane}
due to the JT energy gap between all neigboring sites,  
so dispersion along 
$k_x$ and $k_y$ is 
small.
For classical spins, 
 charge  carriers 
are fairly localized 
 in {\em all three} directions, 
as  neigboring  planes 
have opposite 
spins.
For quantum spins, however, 
inter--plane  hopping 
{\em between  JT--undistorted sites}  (Fig. 1(a))
becomes possible 
by deforming the ground state AFM spins.
This  results in  
 large dispersion  along $k_z$, 
only for 
the spin--dressed conduction quasi--electrons 
close to $\varepsilon$=0 
(Fig. 5(a)). 
As in Fig. 4,
polaronic holes do not deform strongly 
the parallel
background  spins, so
the valence band dispersion in Fig. 5(a),
close to $\varepsilon$=$-E_{JT}$,
is small.
Relaxation
across the large insulator gap is 
 suppressed, 
so photoexcitation 
creates 
non--equilibrium  $e$ and $h$ 
  populations
{\em with two very different chemical potentials 
and spin properties}. 
A critical  
density of 
quasi--electrons 
in the 
dispersive  conduction band 
leads to {\em anisotropic metallic properties}  
and global conductivity.
The 
(FM) 
spin--canting 
responsible for this mobility 
dominates over JT distortion in determining the free energy 
change 
 \cite{Krish}. 
In contrast, holes
 have weak dispersion, small Fermi energy, 
and 
localize 
by relaxing 
 JT  distortions  to gain elastic energy
 \cite{Krish}. 
The differences 
between classical and quantum spins become most pronounced 
as $E_{JT}$   decreases to values 
reasonable for some manganites 
(Figs. 5(b) and 5(d)).
Quantum spin fluctuations can then overcome JT confinement to enhance
delocalization of {\em both} $e$ and $h$ quasi--particles 
in all three directions
and rapidly close the charge energy gap.
 Decreasing $E_{JT}$ 
favors  an  insulator--to--metal  transition
\cite{Ramakrish} 
for quantum spins, which  
 may explain why classical 
spin  calculations 
require critical magnetic fields for CMR 
phase transition much larger  \cite{cant}
than experiment 
 \cite{Krish}. 
Note that the presence of lattice deformations \cite{Hwang,RiniNat} can work cooperatively with the proposed electronic fluctuation mechanism to decrease $E_{JT}$ and further enhance the above effects. 

In summary,
by 
simultaneously tracing 
the {\em fs} dynamics of charge and spin excitations,
we provide solid evidence that 
the properties of 
CMR manganites 
are governed by the coexistence of two 
very different quasi--particles: 
metallic 
 quasi--electrons 
 dressed by quantum spin fluctuations
and JT polarons.
Femtosecond  coherent nonlinear excitation 
controls a  ``sudden'' AFM$\rightarrow$FM  switching 
in the insulating phase, 
coincident with majority/minority carrier 
reversal and 
closing of the 
JT energy gap by quantum spin fluctuations. 

This work was supported by the National Science Foundation 
Contract No. DMR-1055352 (ultrafast laser spectroscopy). The computational studies were supported by the European Union's Seventh Framework Programme (FP7-REGPOT-2012-2013-1)
under grant agreement No. 316165, and by 
the EU Social Fund and National resources 
through the THALES program NANOPHOS.

\section{Supplementary Information} 

\section{Composite Fermion Quasi--particles}

Here we summarize the formalism 
used to describe the dynamical  coupling of spin and charge 
excitations in the strong--coupling insulating limit.
Our observations of 
ultrafast 
electron hopping  correlated with fs spin dynamics 
suggest that this is a key many--body mechanism in 
 manganites.  
Classical  spin 
scenarios 
 assume an adiabatic approximation 
of  electrons  scattering 
off a frozen spin configuration \cite{Krish}. 
The effective inter--atomic 
hopping amplitude 
then decreases with increasing angle between the local 
spins \cite{Dagotto,Krish,anderson}
and electron hopping 
is suppressed by AFM spin alignment. For the CE--AFM 
reference state, this 
results in electronic confinement  
within  one--dimensional FM chains 
with parallel $e_g$ and $t_{2g}$ spins
\cite{Dagotto,Brink}.  
Spin dynamics directly correlated with {\em simultaneous} 
photoelectron femtosecond 
motion is  impossible within the classical spin 
adiabatic approximation.  
We therefore had to turn to non--adiabatic \emph{quantum spin} 
scenarios 
in order to explain our observation of correlated 
simultaneous spin--charge femtosecond dynamics.

Our proposed theory
addresses the following issues:  
 (i) local 
interactions well--exceed the kinetic energy and restrict the 
population of certain atomic configurations, 
(ii) while here 
strong on--site interactions result in an
insulating  ground state,
the spin properties are still determined 
by electron hopping between different atoms \cite{Ramakrish}. 
In equilibrium,
virtual inter--atomic 
electronic fluctuations 
 across the JT insulator energy gap 
lead to a FM exchange coupling \cite{Krish,Ramakrish}. 
During  fs laser excitation, 
coherent $e$--$h$ 
excitations are 
driven across the JT  gap 
and controlled via 
optical field 
Rabi energy and central frequency. 
Here we show that these driven 
fast charge 
fluctuations result in non--equilibrium 
 ``sudden'' FM correlation,
(iii) our experimental results indicate 
that the coherent charge excitations 
bring the system away from equilibrium while 
{\em simultaneously} exciting 
the spin degrees of freedom. 
 This observation 
suggests that non--adiabatic spin dynamics  during
 femtosecond electronic timescales
is key for explaining our experiment, while  
the classical spin adiabatic approximation
assumes independent spin and charge degrees of freedom
\cite{Dagotto,Krish}. 
Below we
introduce composite fermion quasi--particles 
with coupled spin--charge degrees of freedom.

Since the strong interactions are  local,
we start with atomic states that 
diagonalize the  Hund's rule and 
JT
interactions at given atom $i$. 
Each atom can be populated by 0 or 1
mobile ($e_g$) electrons,  since 
 occupancy by two $e_g$ electrons 
(Mn$^{2+}$ configurations) 
is suppressed, e.g.  
by large Hubbard--U 
repulsion. 
The atomic states with a single $e_g$ electron are 
\begin{eqnarray} 
|i \alpha
 M  \rangle 
&=&
\sqrt{\frac{S + M + \frac{1}{2}}{
{2S + 1}}} 
 \ c^{\dag}_{ 
i \alpha \uparrow} \ |i, M-\frac{1}{2} \rangle \nonumber \\
&& + \sqrt{\frac{S - M + \frac{1}{2}}{2S+1}} \  c^{\dag}_{ 
i \alpha \downarrow} \ |i,M+\frac{1}{2} \rangle,
\label{state} 
\end{eqnarray}
where 
$c^{\dag}_{i \alpha \sigma}$ 
adds  a spin--$\sigma$  $e_g$ electron in 
orbital state $\alpha$
and 
$|i  S_z \rangle$, $S_z$=$-S, \cdots, S$, 
are 
 $S$=3/2 local ($t_{2g}$)
spin states for given lattice 
displacement at site $i$. 
The above 
$J$=$S+$1/2, 
$M$=$-J, \cdots, J$ states are characterized by 
the eigenvalues $(J,M)$ of the total mobile 
($e_g$) plus local ($t_{2g}$)
spin.
For $J_H$$\rightarrow$$\infty$, 
the population of 
$J$=$S-$1/2 states is suppressed.
For $M$=$J$=$S+$1/2,  the 
itinerant and local spins are parallel,
as for  classical spins.
The quantum spin dynamics discussed in the main text arises 
mostly from 
$M$=$S-$1/2  atomic states
 Eq.(\ref{state}),
 with mixed spin
due to the 
 off--diagonal
interaction  
$J_H S^{\pm }_{i}$$\cdot$$s^{\mp }_{i}$.

In the limit of strong correlation, 
we describe  local excitations  
in terms of transitions between the above 
atomic many--body states $| i a \rangle$, 
created by the 
``Hubbard operators'' 
$| i a \rangle \langle i b |$. 
On--site excitations that
conserve the total number of electrons are 
created by the operators 
\begin{equation} 
\hat{X}_{i}(\alpha M; \alpha^{\prime} M^\prime)
= | i \alpha M \rangle 
 \langle i \alpha^{\prime} M^\prime | 
\ , \ 
\hat{X}_{i}(m;m^{\prime})
= | i m\rangle 
\langle i m^{\prime}  |.
\label{X-S}  
\end{equation} 
In the limit of 
large Hubbard--U and Hund's rule magnetic exchange 
interactions, the local ($t_{2g}$) spin
z--component  $S_z(i)$ at site $i$ 
is expressed as 
\begin{equation} 
S_z(i)=
\sum_{m=-S}^S m \, \rho_i(m) + 
\sum_{M=-S-\frac{1}{2}}^{S + \frac{1}{2}}    
M \, \frac{ S}{S + \frac{1}{2}}  
\, \sum_{\alpha} 
 \rho^\alpha_i(M),
\label{spin-loc}  
\end{equation} 
where the 
diagonal density matrix elements 
\begin{eqnarray} 
&& \rho_i(m)=
\langle \hat{X}_i(m,m) \rangle = 
\langle | i  m \rangle \langle i  m | \rangle 
\nonumber \\ 
&&
\rho_i^{\alpha}(M)=
\langle \hat{X}_i(\alpha M; \alpha M) \rangle =
\langle | i \alpha M \rangle \langle i \alpha  M | \rangle,
\label{rho-J} 
\end{eqnarray} 
$m$=$-S, \cdots, S$ and 
$M$=$-J, \cdots, J$,
give the populations of the 
empty and singly--occupied 
atomic configurations. 
Their 
equations--of--motion
couple  off--diagonal density matrix elements  
that describe linear superpositions
of quantum states 
in {\em two different} atoms
($e$--$h$ coherence). 
We describe such $e$--$h$ coherence for 
 strong on--site interactions
by first introducing 
 ``composite fermion''  local excitations 
with fixed {\em  total spin} 
$J_z$=$\sigma$/2. These quasi--electron 
charge excitations  
are 
created 
 by 
 Hubbard operators 
that change the  number of electrons 
on a given atom 
by one 
via 
Mn$^{4+}$$\rightarrow$  
Mn$^{3+}$ 
transitions between the  many--body states
that    diagonalize  the strong 
spin, charge, and lattice  
on--site interactions:
\begin{equation} 
\hat{e}^\dag_{\alpha \sigma}(i M)
= | i \alpha M \rangle \langle i, M - \frac{\sigma}{2}| \ , \
M=-J, \cdots, J.
\label{Hubb-def}
\end{equation}
The $e$--$h$ coherence is characterized by 
the off--diagonal density matrix elements 
$\langle 
\hat{e}^\dag_{\alpha^\prime \sigma^\prime}(i^\prime M^\prime)
\, \hat{e}_{\alpha \sigma}(i M)\rangle$, which are 
defined in terms of composite fermions.
Delocalized excitations 
in a periodic 
lattice of atoms located at positions 
$(i,R_i)$, where  $i$ now labels the 
different atoms in a single unit cell 
and $R_i$ 
is the periodic lattice vector
that labels the different unit cells, 
are described 
by   transforming 
to 
k--space using the Bloch theorem: 
 \begin{equation}
\hat{e}^\dag_{ k  \sigma}(i \alpha M)  
= \frac{1}{\sqrt{N}} \sum_{R_i} 
e^{i k R_i} 
\, \hat{e}^\dag_{\alpha \sigma}( i R_i M),
\label{ek-def} 
\end{equation} 
where 
 $N$ is the number of unit cells
and $k$ the wavevector.

Composite fermions
obey the  non--canonical 
 anti--commutation relations
\begin{widetext}  
\begin{eqnarray} 
[
\hat{e}^\dag_{\alpha^\prime \sigma^\prime}(i^\prime M^\prime),
\hat{e}_{\alpha \sigma}(i M) 
]_{+} 
= 
\delta_{i i^\prime} \
\left[
\delta_{M^\prime,M+\frac{\sigma^\prime-\sigma}{2}} 
\ \hat{X}_i(\alpha^\prime M^\prime; \alpha M) 
+ 
\delta_{M^\prime,M} \ \delta_{\alpha,\alpha^\prime} \  
\hat{X}_i(M-\frac{\sigma}{2},
 M^\prime-\frac{\sigma^\prime}{2}) \right].
\label{fermi2}  
\end{eqnarray} 
\end{widetext}
The difference from 
 fermion anti--commutator is often
referred to as  ``kinematic interaction'' and 
 comes from the 
restriction,
due to  on--site interactions
 exceeding the kinetic energy, 
 in the 
populations of individual atoms 
where the electron is 
allowed to hop.
For example, strong 
 Hubbard repulsion 
and Hund's rule interactions 
suppress doubly--occupied 
and $J$=$S-$1/2 
atomic configurations during electron motion.
We  thus project 
the bare electron operators
 onto the subspace 
of the low--energy populated  states  Eq.(\ref{state}):
\begin{equation} 
\hat{e}_{\alpha \sigma}^{\dag}(i)  
= \sum_M 
F_{\sigma}(M) 
\ \hat{e}^\dag_{\alpha \sigma}(i M),
\label{e-proj}
\end{equation} 
where the 
Glebsch--Gordan coefficients
\begin{equation} 
F_{\sigma}(M)=\sqrt{\frac{S + \frac{1}{2} + \sigma M}{2S + 1}}\label{GG}
\end{equation} 
arise from the conservation of the total spin ${\bf J}$.
The projected time--dependent many--body Hamiltonian 
that describes the 
laser--excited system has the form 
\begin{eqnarray} 
&& H(t)=
\sum_i
\sum_{\alpha M} 
E_{i}(\alpha M) \,  \hat{X}_{i}(\alpha M; \alpha M) 
 \nonumber \\
&&
+ \sum_{i} \sum_{m}
 E_i(m) 
\,  \hat{X}_{i}(m,m)
 + H_{hop}(t).
 \label{H(t)} 
\end{eqnarray}
The first two terms diagonalize exactly the 
many--body atomic Hamiltonian that includes 
all onsite interactions, with eigenvalues 
$E_{i}(\alpha M)$ 
(Mn$^{3+}$) 
 and $ E_i(m)$ (Mn$^{4+}$).  
The eigenvalues 
 $E_{i}(\alpha M)$ of states with  
a single $e_g$ electron 
depend on the lattice displacement at site $i$ 
due to electron--phonon interaction 
with the  local vibrational (JT)   modes.
While the above two terms  dominate in the insulating limit, 
 inter--site electron hopping is 
allowed via fast
charge fluctuations, 
virtual  
\cite{Krish,Ramakrish} 
or  
driven by the time--dependent laser E--field: 
\begin{widetext}
\begin{eqnarray} 
&& H_{hop}(t)=-  \sum_{i i^\prime}  
\sum_{\sigma}  
\sum_{\alpha \alpha^\prime} 
V_{\alpha \alpha^\prime}(i- i^\prime) 
 \left[ \cos \left(\frac{\theta_i - \theta_{i^\prime}}{2} \right) 
\, \hat{e}^\dag_{\alpha \sigma}(i) \,
\hat{e}_{\alpha^\prime \sigma}(i^\prime)
+ \sigma \, 
 \sin \left(\frac{\theta_i - \theta_{i^\prime}}{2} \right) 
\, \hat{e}^\dag_{\alpha \sigma}(i) \,
\hat{e}_{\alpha^\prime -\sigma}(i^\prime) \right].
\label{hop}
\end{eqnarray}
\end{widetext}
For tight--binding Hamiltonians, 
the  hopping amplitude between
sites 
$ r_i$ and $r_j$
is modified by the 
laser
(vector potential 
${\bf A}$(t))
as described 
by the 
Peierls substitution
\begin{equation} 
\label{V-def} 
V_{\alpha \alpha^\prime}(j-i)=
t_{\alpha \alpha^\prime}(j-i) 
\exp[-ie {\bf A}(t) \cdot (r_j - r_i) /\hbar c], 
\end{equation} 
where $t_{\alpha \alpha^\prime}$ are the 
tight--binding parameters.
We decompose
this hopping amplitude into equilibrium and laser--induced parts, 
$V_{\alpha \alpha^\prime}(j-i)= 
t_{\alpha \alpha^\prime} + 
 \Delta V_{\alpha \alpha^\prime}(t)$,
where for our typical pump intensities 
\begin{equation} 
\Delta V_{\alpha \alpha^\prime}(i-j) \approx
 d_0(t) \ \frac{ t_{\alpha \alpha^\prime}(i-j)}{\hbar \omega_p}.
\end{equation}
The time--dependence of the  Rabi energy
$d_0$(t)=$e E(t) a$, 
where $a$ 
is the lattice spacing, 
is determined by  
the 
amplitude of the 
laser field
$\propto e^{-t^2/t_p^2}$. 
$\hbar \omega_p$ is the pump 
 central frequency and we consider  $t_{p}$=100fs.
 
In Eq.(\ref{hop}), 
the  spin--canting angles 
$\theta_i$ characterize the reference 
(equilibrium) state 
and define the  spin background 
within the adiabatic approximation
(we assume zero 
polar angles). 
These angles 
describe the 
tilt of the local z--axis at site $i$, 
defined by the direction of the equilibrium spins, 
 with respect to the 
 laboratory z--axis. 
In the calculations presented here,  
we assume AFM ground state, so   
$\theta_i$=0 at spin--$\uparrow$ sites 
and 
$\theta_i$=$\pi$ at spin--$\downarrow$ sites. 
For $\sigma$=$\uparrow$, the first term on the rhs
of Eq.(\ref{hop}) 
 describes 
the usual coherent electron hopping amplitude \cite{anderson,Krish,Dagotto}, 
which is   maximum 
for 
parallel spins $\theta_i$=$\theta_{i^\prime}$.  
For
$\sigma$=$\downarrow$, this  term  describes 
electron hopping accompanied by simultaneous spin--flips 
on both initial and final sites.
The second term  describes the electron--magnon interaction
\cite{Kapet-corr} 
in the strong--coupling limit,  
which
allows an electron to hop 
between two sites with opposite (AFM) spins. 

\section{Quantum Kinetics of spin--charge  coupling: 
equations of motion  
}

In this section  we summarize the 
quantum kinetic 
density matrix equations of motion  
that we use to describe the simultaneous spin and charge 
excitation while the laser pulse interacts with the material 
(coherent temporal regime). 
These equations are derived  by using the 
 time--dependent  Hamiltonian Eq.(\ref{H(t)}). 
The time evolution of the spin--dependent atomic populations 
(diagonal density matrix elements) is described as follows:
\begin{widetext} 
\begin{eqnarray} 
&& 
i \partial_t  \rho^{\alpha}_i(M) 
=
2 \, Im \,
\frac{1}{N} \, \sum_{k^\prime}  
  \sum_{l
\alpha^\prime }
V^{k^\prime}_{\alpha^\prime \alpha}(l-i)   
\sum_{\sigma^\prime=\pm 1 }  
F_{\sigma^\prime}(M) \times 
\nonumber \\
&& 
 \Bigg[ 
 \cos\left(\frac{\theta_{l} - \theta_i}{2} \right)
\langle \hat{e}^{\dag}_{k^\prime \sigma^\prime}(l \alpha^\prime) 
\, 
\hat{e}_{k^\prime \sigma^\prime}(i \alpha M) \rangle
 - 
\sigma^\prime  
 \sin \left(\frac{\theta_{l} - \theta_i}{2}\right)
\, \langle \hat{e}^{\dag}_{k^\prime  -\sigma^\prime}(l \alpha^\prime) 
\, 
\hat{e}_{k^\prime \sigma^\prime}(i \alpha M) \rangle
\Bigg] \label{dm-Xab-1} 
\end{eqnarray}
determines  
the population of atomic many--body configurations with
 a single $e_g$--electron
  and
\begin{eqnarray} 
&& 
\partial_t  \rho_i(m) 
=-
2 \, Im \, 
\frac{1}{N} \sum_{k^\prime}  
\sum_{l \alpha^\prime \alpha}  V^{k^\prime}_{\alpha^\prime \alpha}(l-i) 
\sum_{\sigma^\prime = \pm 1}
F_{\sigma^\prime}(m+\frac{\sigma^\prime}{2}) 
\times 
\nonumber \\
&&
\Bigg[ 
\cos\left(\frac{\theta_{l} - \theta_i}{2} \right) 
\, 
\langle \hat{e}^{\dag}_{k^\prime \sigma^\prime}(l \alpha^\prime) 
\, \hat{e}_{k^\prime \sigma^\prime}(i \alpha, m+\frac{\sigma^\prime}{2}) \rangle 
- \sigma^\prime \,
 \sin\left(\frac{\theta_{l} - \theta_i}{2} \right) 
\, 
 \langle \hat{e}^{\dag}_{k^\prime -\sigma^\prime}(l \alpha^\prime) 
\, \hat{e}_{k^\prime \sigma^\prime}(i \alpha,  m+\frac{\sigma^\prime}{2}) 
\rangle \Bigg]
\label{dm-XM-1} 
\end{eqnarray}
\end{widetext}
determines  
the population of atomic configurations with
 empty $e_g$ orbitals ($t_{2g}$ spin only). In the above equations,  
 we introduced the Fourier--transform of the 
time--dependent hopping amplitude
$V_{\alpha \beta}$, Eq.(\ref{V-def}),  
assuming a periodic system 
with different atoms $i$ and $j$ per unit cell: 
\begin{equation} 
V^k_{\alpha \beta}(i-j) 
=\sum_R V_{\alpha \beta}(R+i-j) \, e^{-i k R}.
\end{equation} 
The above population equations of motion  are {\em exact} in the limit 
$J_{H}, U$$\rightarrow$$\infty$.
They describe the  dynamical  coupling of any given atom  $i$ 
to the 
rest of the lattice,
driven by $H_{hop}(t)$ Eq.(\ref{hop}). This dynamics is 
 characterized  by the time--dependent 
inter--atomic $e$--$h$ coherences 
$ \langle \hat{e}^{\dag}_{k^\prime \pm \sigma^\prime}(l \alpha^\prime) 
\, \hat{e}_{k^\prime \sigma^\prime}(i \alpha, 
 m+\frac{\sigma^\prime}{2}) \rangle$, Fig. 1(b), 
of composite fermion excitations 
(rather than bare--electrons as with 
$e$--$h$ excitations in weakly correlated systems \cite{Chemla,Axt98}). 
These coherences 
characterize  transient superpositions  
of spin--dependent  many--body atomic states  
in the  pair of atoms 
$i$ and $l$ in the unit cell. 
Of main interest  here is the 
laser--driven time--dependent coherence
across the JT insulator gap, between 
 JT--distorted sites and 
undistorted sites.
In  real space, 
the dynamics of inter--atomic coupling is determined  
by the following  (exact) equations of motion:
\begin{widetext} 
\begin{eqnarray} 
&& i \partial_t  \langle \hat{e}^{\dag}_{\beta \bar{\sigma}}(j)
\, \hat{e}_{\alpha \sigma}(i M) \rangle
-\left [\varepsilon_{\alpha \sigma}(i)  
-\varepsilon_{\beta \bar{\sigma}}(j) 
\right] \, \langle \hat{e}^{\dag}_{\beta \bar{\sigma}}(j)
\, \hat{e}_{\alpha \sigma}(i M) \rangle 
\nonumber \\
&& =
\sum_{l \sigma^{\prime}}
\sum_{\alpha^\prime \beta^{\prime}}
V_{
\alpha^\prime \beta^\prime}
(l-j) \, 
\cos\left(\frac{\theta_{l} - \theta_{j}}{2} \right) 
\, 
\langle 
  [ \hat{e}^\dag_{\beta \bar{\sigma}}(j),
\hat{e}_{\beta^\prime \sigma^\prime}(j)
]_+ \, \hat{e}^\dag_{\alpha^{\prime} \sigma^{\prime}}(l) \, 
  \hat{e}_{\alpha \sigma}(i M) \rangle 
\nonumber \\
&&
 -
 \sum_{l \sigma^\prime} 
\sum_{\alpha^\prime \beta^{\prime}}
V_{
\alpha^\prime \beta^\prime}(i-l)  \,
\cos\left(\frac{\theta_{l} - \theta_{i}}{2} \right) 
\, \langle  \hat{e}^{\dag}_{\beta \bar{\sigma}}(j) \,  
\hat{e}_{ \beta^{\prime} \sigma^{\prime}}(l) 
\,  [ \hat{e}^\dag_{\alpha^\prime \sigma^\prime}(i),
\hat{e}_{\alpha \sigma}(i M)
]_+ \, 
\rangle 
\nonumber \\
&&
+  \sum_{l \sigma^{\prime}}
\sum_{\alpha^\prime \beta^{\prime}}
V_{
\alpha^\prime \beta^\prime}
(l-j) \, \sigma^\prime
\, \sin\left(\frac{\theta_{l} - \theta_{j}}{2} \right) 
\, 
\langle 
  [ 
\hat{e}^\dag_{\beta \bar{\sigma}}(j),
\hat{e}_{\beta^\prime -\sigma^\prime}(j)
]_+ \, 
\hat{e}^\dag_{\alpha^{\prime} \sigma^{\prime}}(l) 
\, 
  \hat{e}_{\alpha \sigma}(i M) \rangle 
\nonumber \\
&&
 +
 \sum_{l \sigma^\prime} 
\sum_{\alpha^\prime \beta^{\prime}}
V_{
\alpha^\prime \beta^\prime}(i-l)  \, 
\sigma^\prime \,
\sin\left(\frac{\theta_{l} - \theta_{i}}{2} \right) 
\, 
\langle  \hat{e}^{\dag}_{\beta \bar{\sigma}}(j) \,  
\hat{e}_{ \beta^{\prime}- \sigma^{\prime}}(l) 
\,  [ \hat{e}^\dag_{\alpha^\prime \sigma^\prime}(i),
\hat{e}_{\alpha \sigma}(i M)
]_+
\rangle,  
\label{eom-coh-real} 
\end{eqnarray} 
\end{widetext} 
where we introduced the 
many--body excitation energies 
\begin{equation} 
\varepsilon_{\alpha \sigma}(i) 
=
E_{i}(\alpha M) 
- E_{i}(M - \frac{\sigma}{2}).
\label{e-energ} 
\end{equation}
Eq.(\ref{eom-coh-real}) 
 treats non--adiabatic quantum spin  dynamics
during the  fast electronic motion and  
couples dynamically  spin 
and charge excitations.

Coherence between many--body states 
across the Mott--Hubbard insulator gap
 couples the two Hubbard bands 
 and is at the heart of 
the insulator--to--metal phase transition. 
Here, Eq.(\ref{eom-coh-real})  
was derived in  the limit 
$U,J_H$$\rightarrow$$\infty$, where  
the finite insulator energy gap arises
between  JT--distorted and 
undistorted sites.
The laser excitation drives 
femtosecond non--equilibrium coherence 
across this JT insulator gap.
The differences from the familiar 
 equations of motion for  $e$--$h$ 
coherence 
 in weakly--correlated systems \cite{Chemla,Axt98}
arise from the deviation  
of the {\em composite fermion} anti--commutators
$ [ \hat{e}^\dag_{\alpha^\prime \sigma^\prime}(i),
\hat{e}_{\alpha \sigma}(i M)
]_+$
from fermion behavior (kinematic interaction).
 This
effect of strong local correlations 
couples  two-- and 
four--particle density matrices, leading to a many--body 
hierarchy. 
The spin--dependent composite fermion operators 
$\hat{e}^\dag_{\alpha \sigma}(i M)$, Eq. (\ref{Hubb-def}), 
 not only project--out double--occupancy of any site, 
but also  distinguish between 
many--body atomic populations with different 
{\em total spin}, after 
 diagonalizing exactly the 
magnetic exchange interaction   
that mixes individual spins. 
  $\sigma$
labels the {\em total spin} of the many--body excitation 
$| i M-\frac{\sigma}{2} \rangle$$\rightarrow$
$| i \alpha M>$ created by 
$\hat{e}^\dag_{\alpha \sigma}(i M)$
and  coincides with the bare electron spin
only if we neglect local spin excitation 
during  electronic hopping,  
as in the classical spin adiabatic approximation.  
Eq.(\ref{eom-coh-real})  
couples $\pm \sigma$ 
local excitations in sites $i$ and $j$ with 
different equilibrium spin orientations, 
$\theta_i$$\ne$$\theta_j$, and describes
electron motion 
in AFM systems.

To truncate the hierarchy of equations of motion, 
we  factorize  the four--particle density matrices of 
{\em composite fermions} (rather than bare electrons)  in 
Eq.(\ref{eom-coh-real}) similar to  the 
 Gutzwiller wavefunction approximation in  infinite dimensions.  
In particular, we use the factorization 
\begin{widetext} 
\begin{eqnarray}
&& \langle 
  [ \hat{e}^\dag_{\beta \bar{\sigma}}(j),
\hat{e}_{\beta^\prime \sigma^\prime}(j)
]_+ \, \hat{e}^\dag_{\alpha^{\prime} \sigma^{\prime}}(l) \, 
  \hat{e}_{\alpha \sigma}(i M) \rangle=
\langle 
  [ \hat{e}^\dag_{\beta \bar{\sigma}}(j),
\hat{e}_{\beta^\prime \sigma^\prime}(j)
]_+ 
\rangle 
\langle 
 \hat{e}^\dag_{\alpha^{\prime} \sigma^{\prime}}(l) \, 
  \hat{e}_{\alpha \sigma}(i M) \rangle,
\label{factor} 
\end{eqnarray} 
where $j$$\ne$$l,i$, 
which after 
using total spin conservation
transforms 
Eq.(\ref{eom-coh-real}) to  
 \begin{eqnarray} 
&& i \partial_t  \langle \hat{e}^{\dag}_{\beta \bar{\sigma}}(j)
\, \hat{e}_{\alpha \sigma}(i M) \rangle  
-\left [\varepsilon_{\alpha \sigma}(i)  
-\varepsilon_{\beta \bar{\sigma}}(j)  
\right] \, \langle \hat{e}^{\dag}_{\beta \bar{\sigma}}(j)
\, \hat{e}_{\alpha \sigma}(i M) \rangle  
\nonumber \\
&&
=\sum_{l} 
\sum_{\alpha^\prime \beta^{\prime}}
V_{
\alpha^\prime \beta^\prime}
(l-j) 
\ \langle  [ \hat{e}^\dag_{\beta \bar{\sigma}}(j),
\hat{e}_{\beta^\prime \bar{\sigma}}(j)
]_+ \rangle  
\ \langle 
 \left[ 
\cos\left(\frac{\theta_{l} - \theta_{j}}{2} \right) 
\hat{e}^\dag_{\alpha^{\prime} \bar{\sigma}}(l) 
- \bar{\sigma} 
 \sin\left(\frac{\theta_{l} - \theta_{j}}{2} \right)  
 \hat{e}^\dag_{\alpha^{\prime} -\bar{\sigma}}(l) \, 
  \right]  \hat{e}_{\alpha \sigma}(i M) \rangle
\nonumber \\
&&
 -  
 \sum_{l} 
\sum_{\alpha^\prime \beta^{\prime}}
V_{
\alpha^\prime \beta^\prime}(i-l) \ 
\langle  [ \hat{e}^\dag_{\alpha^\prime \sigma}(i),
\hat{e}_{\alpha \sigma}(i M)
]_+ \rangle \
\langle 
\hat{e}^{\dag}_{\beta \bar{\sigma}}(j) 
\left[ \cos\left(\frac{\theta_{l} - \theta_{i}}{2} \right) 
 \hat{e}_{ \beta^{\prime} \sigma}(l) 
- \sigma 
\sin\left(\frac{\theta_{l} - \theta_{i}}{2} \right) 
\hat{e}_{ \beta^{\prime} - \sigma}(l) 
\right] \rangle. 
\label{HF-coh} 
\end{eqnarray}
In a periodic system, we can take advantage of the Bloch theorem 
and  Fourier--transform 
the above equation to $k$--space
by using Eq.(\ref{ek-def}). 
We consider for simplicity two neigboring 
atoms in the  unit cell, 
$i$$\ne$$j$,
and neglect the energy dispersion
$V^{k}_{
\alpha \beta}
(0)$ in the insulating limit.  
We  then obtain   in k--space
\begin{eqnarray} 
&& i \partial_t  \langle \hat{e}^{\dag}_{k \bar{\sigma}}(j \beta)
\, \hat{e}_{k \sigma}(i \alpha M) \rangle
-\left [\varepsilon_{\alpha \sigma}(i)  
-\varepsilon_{\beta \bar{\sigma}}(j) 
\right] \, \langle \hat{e}^{\dag}_{k \bar{\sigma}}(j \beta)
\, \hat{e}_{k \sigma}(i \alpha M) \rangle  
 \nonumber \\
&&
=\left[\delta_{\sigma,\bar{\sigma}}
 \cos\left(\frac{\theta_{i} - \theta_{j}}{2} \right) 
+ \sigma \delta_{\sigma,-\bar{\sigma}}
\, \sin\left(\frac{\theta_{i} - \theta_{j}}{2} \right) 
\right] \times 
\nonumber \\
&&
\sum_{\alpha^\prime \beta^{\prime}}
V^{k}_{
\alpha^\prime \beta^\prime}
(i-j)  
\, 
\left[ 
\langle
  [ \hat{e}^\dag_{\beta \bar{\sigma}}(j),
\hat{e}_{\beta^\prime \bar{\sigma}}(j)
]_+ \rangle \, \langle \hat{e}^\dag_{k \sigma}(i \alpha^{\prime}) \, 
  \hat{e}_{k \sigma}(i \alpha M) \rangle 
-  \langle [ \hat{e}^\dag_{\alpha^\prime \sigma}(i),
\hat{e}_{\alpha \sigma}(i M)
]_+ \rangle \, 
\langle 
 \hat{e}^{\dag}_{k \sigma}(j \beta) \,  
\hat{e}_{ k \sigma}(j \beta^{\prime}) 
\rangle 
 \right]. 
\label{dm-coh-k} 
\end{eqnarray} 
In the  above equation, local correlations described by the 
composite fermion anti--commutators modify  
the  coherence and populations of the delocalized carriers.
In the deep insulating limit
$t \ll E_{JT}$, 
we further simplify the problem by 
neglecting 
long--range coherence 
between 
different unit cells
and approximate 
\begin{eqnarray} 
&& \langle \hat{e}^\dag_{k \bar{\sigma}}(i \beta) 
  \hat{e}_{k \sigma}(i \alpha M) \rangle 
\approx \delta_{\sigma \bar{\sigma}} \, F_\sigma(M) 
\, \rho_i^{\beta \alpha}(M)
= \frac{1}{N} \sum_{k^\prime} 
\langle \hat{e}^\dag_{k^\prime \bar{\sigma}}(i \beta) 
  \hat{e}_{k^\prime \sigma}(i \alpha M) \rangle.
\end{eqnarray} 
After using Eqs.(\ref{fermi2}) 
and (\ref{e-proj}) 
to 
express the composite 
fermion  anti--commutators in terms of the local density matrix,
\begin{eqnarray} 
&& \langle [
\hat{e}^\dag_{\alpha \sigma^\prime}(i^\prime),
\hat{e}_{\alpha \sigma}(i) 
]_{+} \rangle=     
 \delta_{i i^\prime} 
\delta_{\sigma \sigma^\prime} 
\left[ \sum_{M=-S-\frac{1}{2}}^{S+\frac{1}{2}} 
F_\sigma^2(M) \
\rho_i^{\alpha}(M)
+  \sum_{m=-S}^S 
F_\sigma^2(m+\frac{\sigma}{2}) \ 
\rho_{i}(m) \right],
\label{electr-anticomm}
\end{eqnarray} 
Eq.(\ref{dm-coh-k}) reduces 
to 
\begin{eqnarray} 
&& i \partial_t  \langle \hat{e}^{\dag}_{k \bar{\sigma}}(j \beta)
\, \hat{e}_{k \sigma}(i \alpha M) \rangle
-\left [\varepsilon_{\alpha \sigma}(i)  
-\varepsilon_{\beta \bar{\sigma}}(j) 
\right] \, \langle \hat{e}^{\dag}_{k \bar{\sigma}}(j \beta)
\, \hat{e}_{k \sigma}(i \alpha M) \rangle  
=  V^k_{
\alpha \beta}
(i-j) \, F_\sigma(M)  \times 
\nonumber \\
&&  
\left[\delta_{\sigma,\bar{\sigma}}
 \cos\left(\frac{\theta_{i} - \theta_{j}}{2} \right) 
+ \sigma \delta_{\sigma,-\bar{\sigma}}
\, \sin\left(\frac{\theta_{i} - \theta_{j}}{2} \right) 
\right]
\, 
 \sum_{M^\prime} 
F^2_{\bar{\sigma}}(M^\prime) \, 
\Bigg[ 
\rho^{\alpha}_i(M) 
\, \rho_j(M^\prime - \frac{\bar{\sigma}}{2}) 
-
\rho_i(M - \frac{\sigma}{2}) \,  
\rho^{\beta}_j(M^\prime) \Bigg]. 
\label{eom-coh-Fourier} 
\end{eqnarray} 
\end{widetext} 
The population product
on the rhs 
describes  population--inversion 
nonlinearity and nonlinear saturation. 
Eqs.(\ref{dm-Xab-1}), 
(\ref{dm-XM-1}),
and (\ref{eom-coh-Fourier}) provide a closed 
system of equations of motion  used  to 
obtain the time--dependent results of Fig. 4. 
Eq.(\ref{eom-coh-Fourier})
 treats short--range coherence between 
nearest--neigbor atoms
with different JT disotrions, 
which  
form ``quantum dimers'' due to 
coherent  coupling by the laser field
across $E_{JT}$. This provides a first 
correction 
to the atomic limit. 
The quasi--equilibrium state 
of the system 
can be  obtained by solving the above equations 
in the adiabatic limit,
by setting 
 $\partial_t$=0. 
For classical spins, 
this recovers  the equilibrium results of Ref. \cite{Krish,Ramakrish},
where a population--dependent FM exchange interaction was obtained 
from virtual charge fluctuations between 
neigboring atoms with different JT distortions.

\section{Composite--Fermion Quasi--particle 
Energy Dispersions for  CE--AFM insulator
reference state}

In this section,
we summarize the calculation
of the  composite--fermion  excitation energy  dispersion 
for a CE--AFM/CO/OO   reference  state with 
coupled spin, charge, lattice, and orbital long--range orders.
For this we 
consider the coherent coupling of the Mn$^{4+}$$\rightarrow$Mn$^{3+}$
 excitations 
Eq.(\ref{Hubb-def})
at different lattice 
sites and  address
 the coupling of different AFM--coupled chains and  planes 
(Fig. 1(a)) 
via 
quantum spin canting. The latter 
      occurs simultaneously 
with electron delocalization along the FM chains. 
We recall that, for classical spins and 
$J_H \rightarrow \infty$,  
different  chains 
are uncoupled 
in the absence of spin canting,
leading to 1D electron confinement 
 \cite{Dagotto,Brink}.

We express the quasi--particle excitations
$\hat{e}_{n k }$
 with momentum $k$ 
in the form  
\begin{equation} 
\hat{e}_{n k } =
 \sum_{i \alpha \sigma} 
u_{k \sigma}^{n}(i \alpha)
\frac{\hat{e}_{k \sigma}(i \alpha)}{\sqrt{n_{\alpha \sigma}(i)}},
\label{expand} 
\end{equation} 
where we introduced the normalization constant
\begin{equation} 
n_{\alpha \sigma}(i)
= 
\langle [\hat{e}^\dag_{k \sigma}(i \alpha), 
\hat{e}_{k \sigma}(i \alpha)]_+ \rangle
= 
 \langle [
\hat{e}^\dag_{\alpha \sigma}(i),
\hat{e}_{\alpha \sigma}(i) 
]_{+} \rangle    
\label{electr-norm}
\end{equation}
determined by Eq.(\ref{electr-anticomm}), 
with average value taken in the  reference state. 
In the 
CE--AFM  background,
 all spins point parallel to the local z--axes
(spin populations $M$=$S$+1/2 
and $m$=S only).
In the calculations of Fig. 5, 
these local axes  point either parallel or 
anti--parallel to the laboratory 
z--axis.
Our 
CE/CO/OO
$x$=1/2 periodic 
reference state is 
characterized by a  three--dimensional unit cell 
of 16 sites
in two adjacent planes, with 
two antiferromagnetically--ordered 
 zig--zag FM chains per plane, 
with four sites per chain,  
alternating 
Mn$^{3+}$ and 
Mn$^{4+}$ atoms,
and neigboring planes with opposite spins but identical 
 charge/orbital  configurations \cite{Dagotto}. 
For such configuration 
we obtain from Eq.(\ref{electr-anticomm}) after using the 
Glebsch--Gordan coefficients 
\begin{equation} 
n_{\alpha \uparrow}(i)
= 
\rho_i^{\alpha}(S+\frac{1}{2})
+ 
\rho_{i}(S) \ , \ 
n_{\alpha \downarrow}(i)
= 
\frac{\rho_{i}(S)
}{\sqrt{2S + 1}}.
\label{electr-norm}
\end{equation} 
$\sigma$=$\uparrow$ corresponds to quasi--electron spin parallel to the 
reference  state local spin,
which is  always the case in the classical spin limit. 
$n_{\alpha \uparrow}(i)$ is a phase space filling contribution 
that prohibits double occupancy, similar to the 
Gutzwiller wavefunction approximation and
slave boson calculations \cite{Brink}.  
$n_{\alpha \uparrow}(i)$=1 similar to Ref. \cite{Dagotto} 
if we neglect double--occupancy.
More importantly, 
here $\sigma$=$\downarrow$ is also allowed 
and describes the possibility 
that the quasi--electron spin may be pointing anti--parallel to the ground state spin, 
as in the quantum  states  Eq.(\ref{state}). 

The laser field excites $e$--$h$ quasi--particles 
on top of the equilibrium reference 
state as described in the 
 previous section. 
We  thus seek here $e$--$h$ eigenmodes 
of the adiabatic Hamiltonian:
\begin{equation} 
i \partial_t  \langle \hat{e}^{\dag}_{n k}
\hat{e}_{m k}
\rangle 
= \left( \omega_{mk} -
\omega_{nk} \right)
 \langle \hat{e}^{\dag}_{n k}
\hat{e}_{m k}
\rangle,
\end{equation} 
where the transformed density matrix 
$ \langle \hat{e}^{\dag}_{n k}
\hat{e}_{m k}
\rangle$ is 
defined by using Eqs.(\ref{expand})
and (\ref{ek-def}).
The time evolution 
Eq.(\ref{HF-coh}) implies 
quasi--particle normal modes 
defined by the following eigenvalue equation: 
\begin{widetext} 
\begin{eqnarray} 
&& \omega_{k m}  u_{k \sigma}^{m}(j \beta)
= \varepsilon_{\beta \sigma}(j)  
\, u_{k \sigma}^{m}(j \beta)
- 
 \sum_{l \alpha }  t^{k}_{\alpha \beta}(l-j)\,  \sqrt{n_{\beta \sigma}(j)} 
\, \sqrt{n_{\alpha \sigma}(l)}
\,
\cos\left(\frac{\theta_{l} - \theta_{j}}{2} \right)
 u_{k \sigma}^{m}(l \alpha)
 \nonumber \\
&&
+\sigma  
 \sum_{l \alpha }   t^{k}_{\alpha \beta}(l-j) \, 
\sqrt{n_{\beta \sigma}(j)} 
\, \sqrt{n_{\alpha -\sigma}(l)} \, 
\sin\left(\frac{\theta_{l} - \theta_{j}}{2} \right) 
 u_{k -\sigma}^{m}(l \alpha),
\label{u-eigen}  
\end{eqnarray} 
\end{widetext} 
where $m$ labels the different quasi--particle branches. 
$\sigma$=+1(-1) means quasi--particle spin parallel (anti--parallel) 
to 
the reference state local spin.
 In the classical
 spin limit, 
$n_{\alpha \downarrow}(i)$=0 and only 
the first term on the rhs of the above 
equation survives, which 
describes energy bands 
due to  coherent hopping 
 amplitude $\propto t^{k}_{\alpha \beta}(l-j)
\cos\left(\frac{\theta_{l} - \theta_{j}}{2} \right)$ \cite{anderson,Krish,Dagotto}.
As in the slave boson \cite{Brink} and Gutzwiller wavefunction 
infinite dimensional 
approximations, 
the local factors 
$\sqrt{n_{\alpha \sigma}(l)}$, 
where here $\sigma$=$\uparrow$,$\downarrow$, 
 modify 
these electron hopping amplitudes
due to  both Hubbard--U 
and magnetic exchange strong local interactions. 
Quantum spins introduce an additional coupling 
between 
$u_{k \uparrow}$ and 
$u_{k \downarrow}$, which breaks degeneracies 
and enhances electron delocalization.
As discussed in the main text,
these local quantum--spin--canting strong correlations
significanly  affect the insulator energy gap and energy band 
dispersions, 
favoring metallic behavior and FM correlations
that strongly depend on $E_{JT}$.
%


\begin{thebibliography}{99}
\bibitem{note1} Our magnetic field is at 
		least one order of magnitude smaller than the critical 
		field needed for CMR transition.
\bibitem{note}  The commonly--used adiabatic approximation 
			assumes  spin dynamics slower than electron dynamics, 
			in which case an effective spin Hamiltonian 
			with spin--exchange interaction mediated by fast electrons
			has meaning.
			It does not apply to the quantum spin--canting 
			dynamics here 
where inter--atomic $e_g$--electron 
			hopping simultaneously flip the $t_{2g}$--electron localized spins.
\bibitem{unit} 
		Our unit cell has 16 sites with alternating 
			Mn$^{3+}$(JT--distorted, polaronic state,  
			orbital points along the chain) and 
			Mn$^{4+}$(undistorted, two degenerate orbitals) 
			atoms,  4 sites per zig--zag chain, 
			two AFM chains per plane,  
			and two AFM planes with identical charge/lattice/orbital
			configurations (Fig. 1(a)). 
			Spin--canting angle $\theta$=0.
\bibitem{cant}
While a ground state spin--canting angle $\theta$
also leads to electron mobility along $k_z$,  
even for classical adiabatic spins, 
a $k_z$--dispersion 
similar to Fig. 5 
($S$=3/2 quantum spins with $\theta$=0)  
requires a
very large $\theta$ 
of few tens of degrees ($\theta$=90$^{o}$ 
corresponds to full FM order without  spin dynamics).
Such spin--canting
requires a very large external magnetic field, much larger 
than in 
CMR experiments. Quantum spin fluctuations, 
which we control here via laser--induced fs $e$--$h$ coherence, 
can explain this discrepancy and greatly enhance the 
sensitivity of fs AFM$\rightarrow$FM phase transitions
far from equilibrium.
\end{thebibliography}
\end{document}